\documentclass[twocolumn]{aastex62}
\usepackage{bm}
\usepackage{fourier}
\usepackage{wasysym}
\usepackage{multirow}

\usepackage[caption=false]{subfig}

\usepackage{hyperref}
\usepackage{natbib}
\newcommand{\vrad}{v_{\rm r}}
\newcommand{\rproj}{r_{\rm proj}}
\newcommand{\Rsub}{R_{\rm sub}}

\shortauthors{Miyauchi \& Kishimoto}
\shorttitle{Three-dimensional distribution of the gas clouds in NGC1068}

\begin{document}
\title{
Velocity-inverted three-dimensional distribution of the gas clouds in the Type 2 AGN NGC1068
}

\correspondingauthor{Makoto Kishimoto}
\email{mak@cc.kyoto-su.ac.jp}

\author{Ryuji Miyauchi}
\affiliation{Department of Astrophysics \& Atmospheric Sciences, Kyoto Sangyo University, Kamigamo-motoyama, Kita-ku, Kyoto 603-8555, Japan}

\author{Makoto Kishimoto}
\affiliation{Department of Astrophysics \& Atmospheric Sciences, Kyoto Sangyo University, Kamigamo-motoyama, Kita-ku, Kyoto 603-8555, Japan}


\begin{abstract}

Spatially-resolved velocity maps at high resolutions of 1-10 pc are becoming available for many nearby AGNs in both optical/infrared atomic emission lines and sub-mm molecular lines. For the former, it has been known that a linear relationship appears to exist between the velocity of the ionized gas clouds and the distance from the nucleus in the inner $\sim$100 pc region, where these clouds are outflowing. Here we demonstrate that, in such a case, we can actually derive the three-dimensional (3D) geometrical distribution of the clouds directly from the velocity map. Revisiting such a velocity map taken by HST for the prototypical Type 2 AGN NGC1068, we implement the visualization of the 3D distribution derived from the map, and show that this inner narrow-line region has indeed a hollow-cone structure, consistent with previous modeling results. Quite possibly, this is the outer extended part of the polar elongated dusty material seen in the recent mid-IR interferometry at pc scale. Conversely, the latter small-scale geometry is inferred to have a hollow-cone outflowing structure as the inward extension of the derived 3D distribution above. The AGN obscuring "torus" is argued to be the inner optically thick part of this hollow-cone outflow, and its shadowed side would probably be associated with the molecular outflow seen in certain sub-mm lines. We discuss the nature of the linear velocity field, which could be from an episodic acceleration that had occurred $\sim$10$^5$ years ago. 

\end{abstract}


\section{Introduction}


Studies of material distribution and its kinematics in AGN is quite central for understanding the physics of accretion and ejection phenomena around supermassive black holes. At 100 pc scales, the kinematics of the inner narrow-line region (NLR) have been spatially resolved by Hubble Space Telescope (HST) and subsequent ground-based integral-field-unit (IFU) instruments  (e.g. \citealt{Hutchings98,Crenshaw00,Cecil02,MullerSanchez11}). This is now becoming even more important to the inner pc and sub-pc scale structure studies, since recent mid-infrared (mid-IR) interferometry has shown that even at a pc scale, the structure is elongated toward the polar, narrow-line-region direction, rather than an equatorial direction (\citealt{Hoenig12,Hoenig13,Tristram14,LopezGonzaga14,LopezGonzaga16}). This polar elongation is interpreted to be due to a polar dusty outflow (\citealt{Hoenig12,Hoenig13}). The mid-IR polar elongation is also seen at scales slightly larger than pc (\citealt{Asmus16,Asmus19}). It is quite likely that this pc scale structure is closely related to the polar structure at 100 pc scale, and conversely, the 100 pc scale structure and kinematics are quite important and constraining on such a small scale.


Furthermore, in sub-mm, high resolution kinematics are now becoming available from ALMA data (e.g. \citealt{GarciaBurillo16,Gallimore16, Imanishi18,Impellizzeri19,GarciaBurillo19}).  The relation between the kinematics of the cold molecular gas and the warm/hot ionized gas has been investigated with matched spatial resolutions (see \citealt{Hoenig19} and references therein).  


The spatially-resolved velocity maps of the NLR have shown that, in the innermost 100 pc (or $\sim$1 arcsec), there seems to be a linear relationship between the radial velocity and the projected distance from the nucleus (\citealt{Crenshaw00,Das06,Das07}). This has been interpreted in such a way that the velocity $v$ of the gas clouds is proportional to the distance $r$ from the nucleus, i.e. $v = k r$ in the three dimensional (3D) space, with this inner region being called an 'acceleration region'. Furthermore, based on the radial velocities from the two-dimensional (2D) spectra at different slit locations, it has been inferred that the spatial distribution of the line emission in 3D has a hollow cone structure. Therefore, in these previous studies, modeling has been done in such a way that the most likely parameters, such as opening angle and inclination, have been determined {\it assuming} a hollow-cone geometry.


Based on these studies, here we demonstrate that the 3D distribution can rather directly be derived from the data, if the linear (or some power-law form) velocity field is assumed. Such a directly derived, empirical 3D distribution would uniquely be beneficial to the inner structure studies. Revisiting the data for the prototypical Type 2 AGN NGC1068 taken with HST, we re-construct and visualize such a 3D distribution directly from the data, and study the structure in detail. We further discuss implications of the derived distribution.

The systemic velocity of NGC1068 is taken as 1148 km/s \citep{Brinks97}, and the distance 14.4 Mpc \citep{Bland-Hawthorn97} where 1" corresponds to 70 pc.


\begin{figure}[t]
\vspace{-1mm}%
\includegraphics[width=9.5cm]{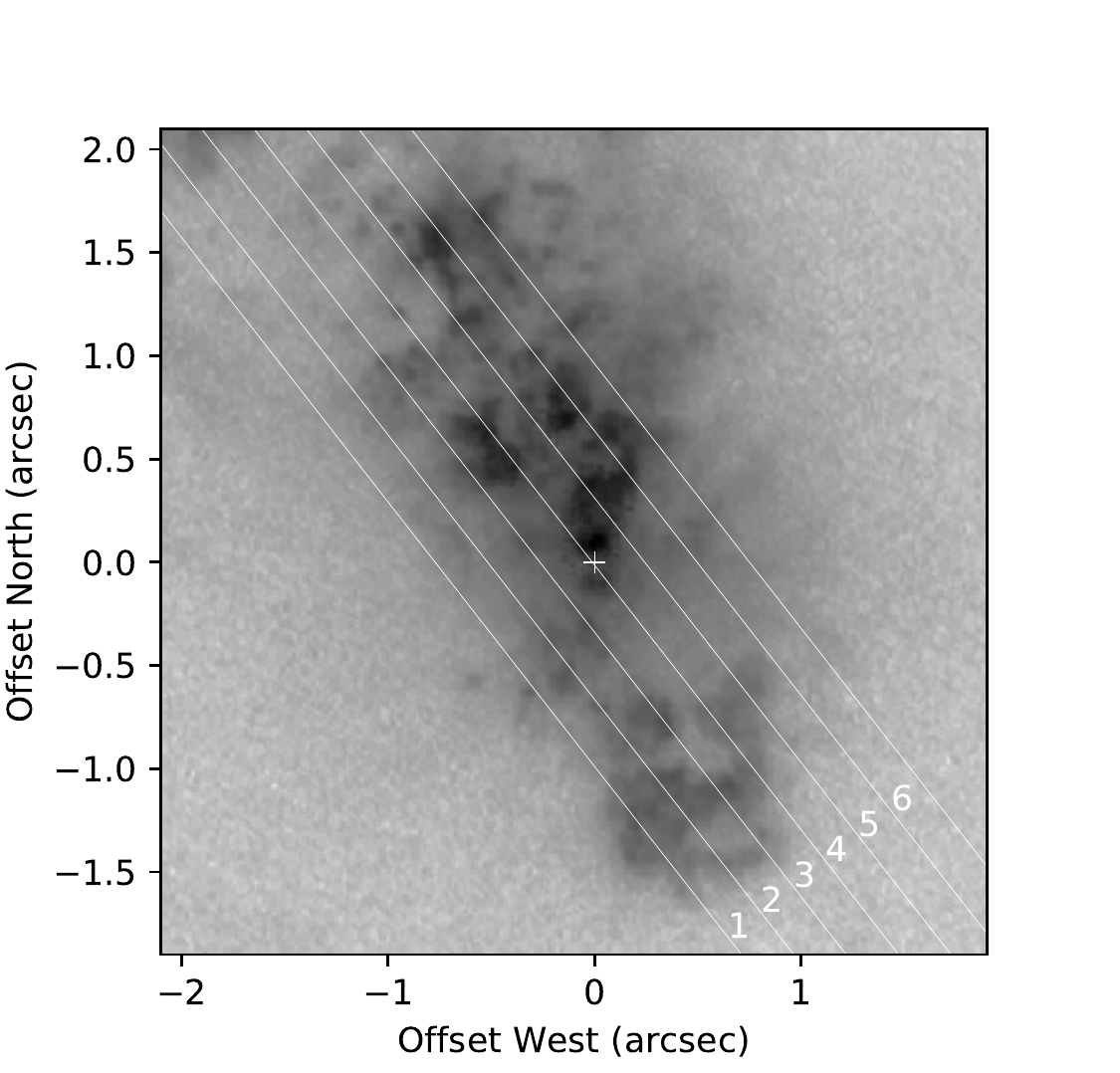}
\caption{Slit positions of STIS observations for NGC1068 (\citealt{Cecil02}) overlaid on the HST/FOC [OIII] image in gray scale (\citealt{Macchetto94}). The STIS slit direction was at $38^{\circ}$ east of north. The position of the hidden nucleus is indicated as a plus sign.}
\label{fig_slitpos}
\end{figure}

\section{Archival data}

\subsection{Data}

We use the data for NGC1068 from the HST archive, taken by \cite{Cecil02} with HST/STIS using the medium resolution grating G430M centered at [OIII] and H$\beta$ line wavelengths (Table~\ref{tab_data}). They were taken at 6 slit positions as shown in Figure~\ref{fig_slitpos}. The data at the 5th slit position were taken redundantly, and the two sets were used to spatially align the data taken a year apart. The slit width was set to be $0.2"$, and the resulting spectral resolution at $\sim$5000 \AA\ is $\sim$5000, corresponding to a kinematical resolution of 60 km/s (STIS Instrument Handbook, chapter 13). Since on-chip binning was done, the spectral sampling of the data is $\sim$0.55 \AA\ per pixel. The spatial sampling is $0.051"$ per pixel (while the HST's spatial resolution at $\sim$5000\AA\ is $\sim$$0.05"$).

\begin{table}[ht!]
\centering  
 \caption{Archival data of STIS long-slit observations for NGC1068 (\citealt{Cecil02})}
 \begin{tabular}{l c c c l}
  \hline \hline
  Name & Observing & Grating & Spectral Range & position\\
       & Date      &         & (\AA)          & \\
  \hline
  o56502010 & 1999 Oct 2  & G430M & 4818-5104 & 1\\
  o56502020 & 1999 Oct 2  & G430M & 4818-5104 & 2\\
  o56502030 & 1999 Oct 2  & G430M & 4818-5104 & 3\\
  o56502040 & 1999 Oct 2  & G430M & 4818-5104 & 4\\
  o56502050 & 1999 Oct 2  & G430M & 4818-5104 & 5\\
  o56503010 & 2000 Sep 22 & G430M & 4818-5104 & 5$^{'}$\\
  o56503020 & 2000 Sep 22 & G430M & 4818-5104 & 6\\
  \hline
 \end{tabular}
 \label{tab_data}
\end{table}

\subsection{Processing}

The data have been cosmic-ray rejected firstly by the HST pipeline, and then by using la-cosmic (\citealt{vanDokkum01}). Remaining hot pixels and negative pixels were filled with the median of adjacent 
pixels. The rest of the processing steps in the pipeline has been implemented using {\it stistools} (ver.1.1). Then, a linear continuum was subtracted using the continuum windows at both sides of the [OIII]$\lambda$4959+5007 line wavelengths. Then the contribution of the [OIII]$\lambda$4959 line was removed by shifting and scaling the [OIII]$\lambda$5007 line.

\begin{figure}[t!]
\includegraphics[width=10cm]{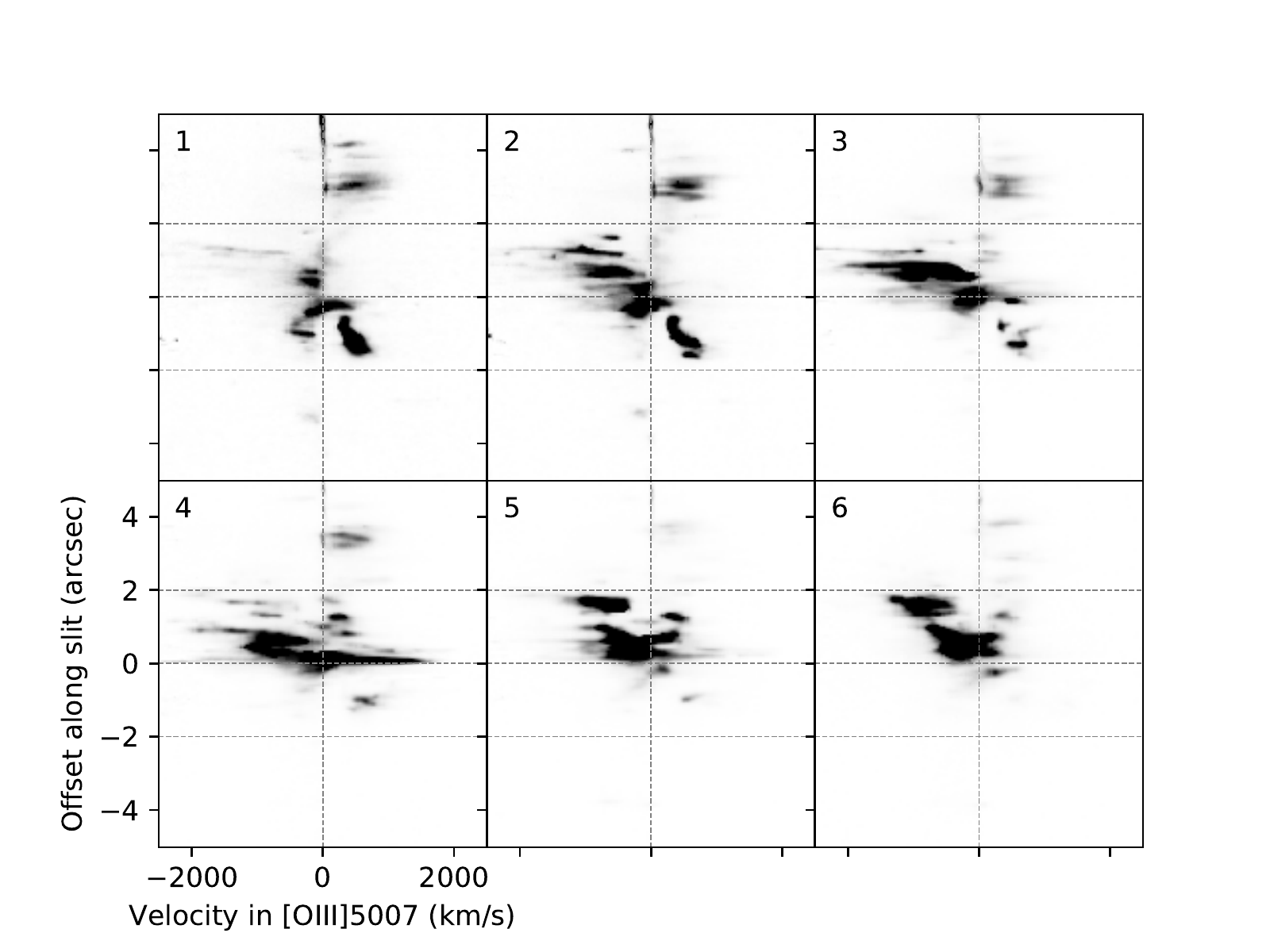}
\caption{HST/STIS 2D spectra at six slit positions as indicated in Fig.\ref{fig_slitpos} with the corresponding position number. The horizontal axis is the dispersion direction, while the vertical axis is along the spatial direction.}
\label{fig_2dspec}
\end{figure}

The systemic velocity of NGC1068 is taken as 1148 km/s (\citealt{Brinks97}), and we have extracted the data over the velocity range of $\pm$2500 km/s of the [OIII]$\lambda$5007. The resulting six 2D spectra are shown in Figure~\ref{fig_2dspec}. These six slit data were assembled to form a data cube of the 2D spatial axes and wavelength axis. Then we interpolated the data cube along the direction perpendicular to the slit direction, in order to have a matched spatial sampling in the two spatial directions (0.05" $\times$ 0.05").

The origin of the spatial coordinates is taken to be the supposed position of the central black hole. This position has been determined as the center of the centro-symmetric polarization pattern in the HST image (\citealt{Capetti95}; \citealt{Kishimoto99}), and it is about $0.1"$ south from the UV/optical continuum peak (see Table~2 in \citealt{Kishimoto99} for details). Below we focus on the central region of $\pm$2" along the slit direction (see more below about this limit).

\begin{figure}[b]
\centering
\includegraphics[width=9cm]{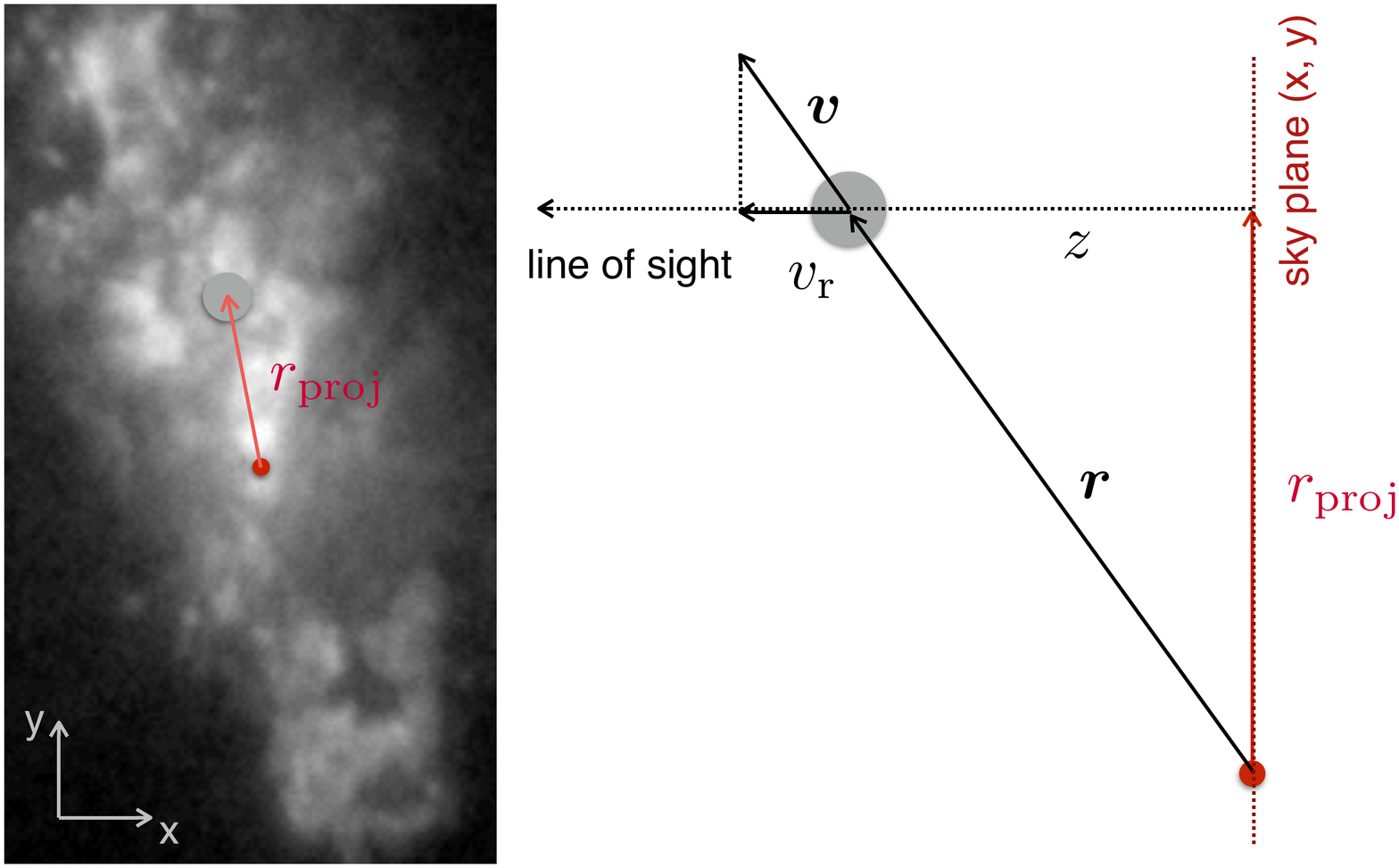}
\caption{Illustration for the variables and observables of a spatially resolved cloud, used in the velocity inversion process for reconstructing the 3D distribution.}
\label{fig_vel_invert}
\end{figure}

\begin{figure*}[ht!]
    \centering
    \subfloat[Volumetric visualization of the 3D flux distribution.]{{\includegraphics[width=0.5\textwidth]{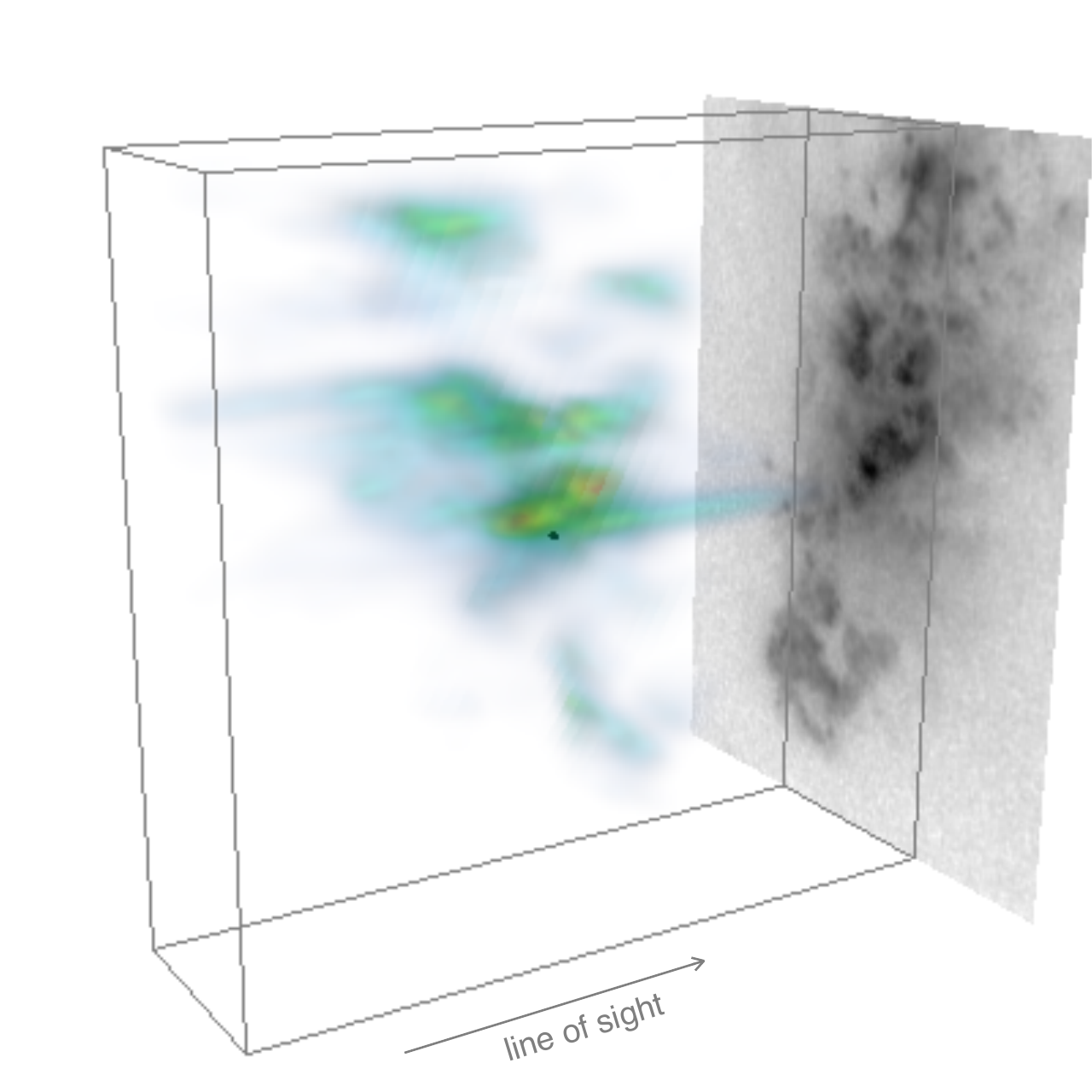} }}%
    \subfloat[Slices of the same 3D distribution in the $x$-$y$ and $y$-$z$ planes through the black hole position.]{{\includegraphics[width=0.5\textwidth]{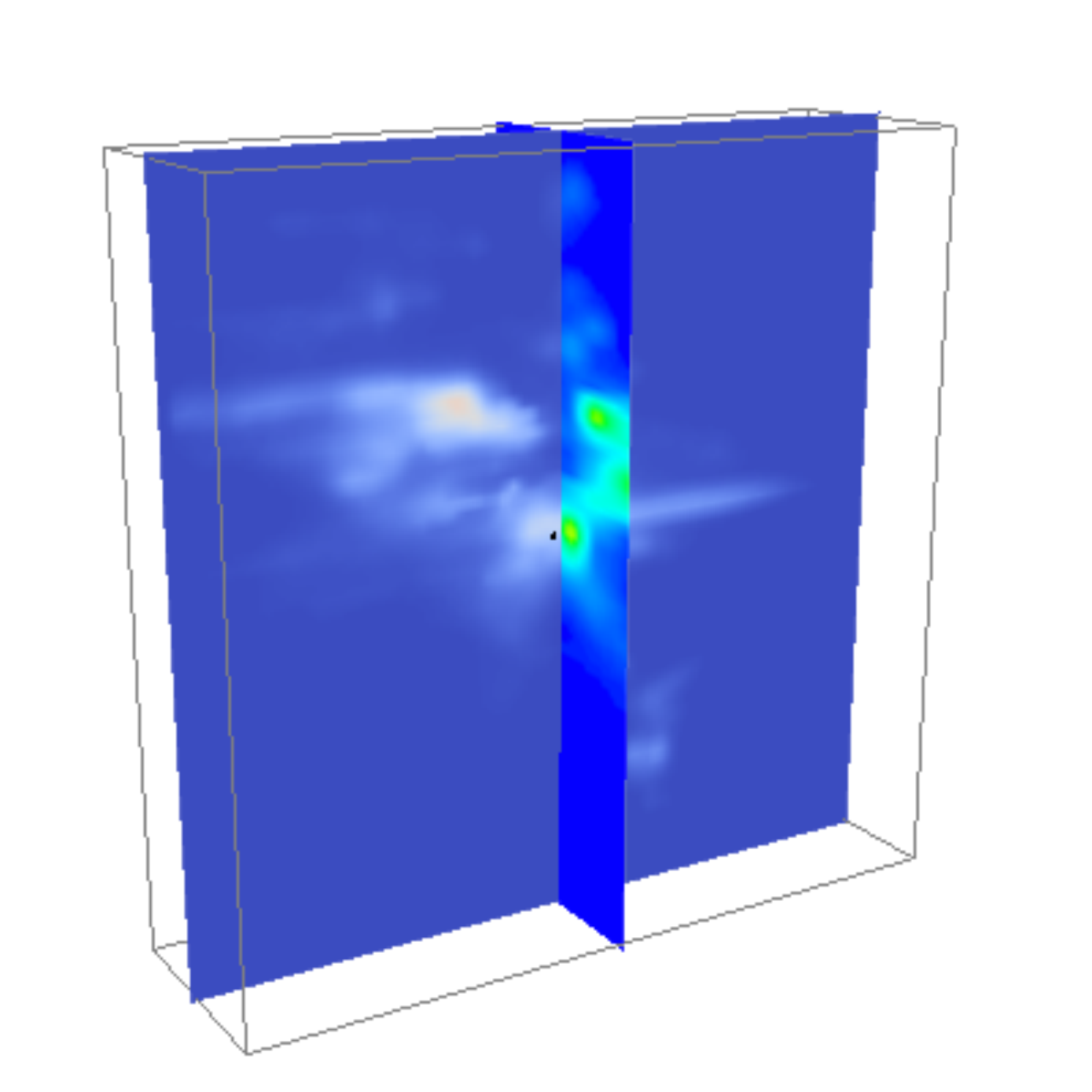} }}%
    \caption{Reconstructed distribution of the [OIII]-line flux in 3D, assuming the linear velocity field $v=kr$ (where $k$ is 1.04$\times$10$^3$ km/s/arcsec or 14.8 km/s/pc; see text). The position of the black hole is indicated as a black point at the center. The reconstructed box scale is 1.2"$\times$4", or $\sim$84$\times$280 pc, parallel to the sky plane, and 280 pc along the line of sight. HST/FOC OIII image is shown for reference as the 2D flux distribution projected on to the sky plane.}
    \label{fig_3D}
\end{figure*}

\begin{figure*}[t]%
    \centering
    \subfloat[slices parallel to sky plane]{{\includegraphics[width=0.5\textwidth]{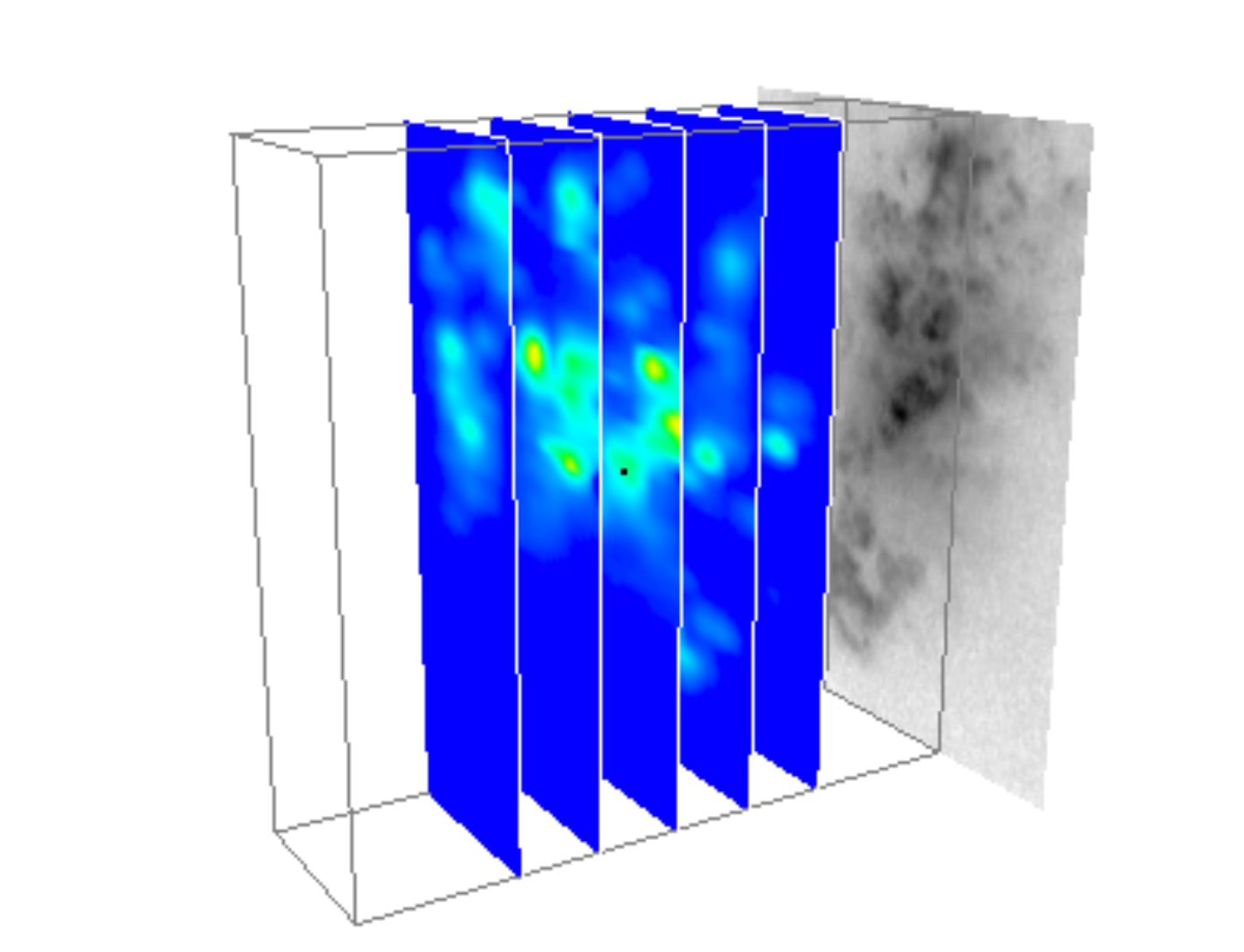} }}%
    %
    \subfloat[slices along line of sight]{{\includegraphics[width=0.5\textwidth]{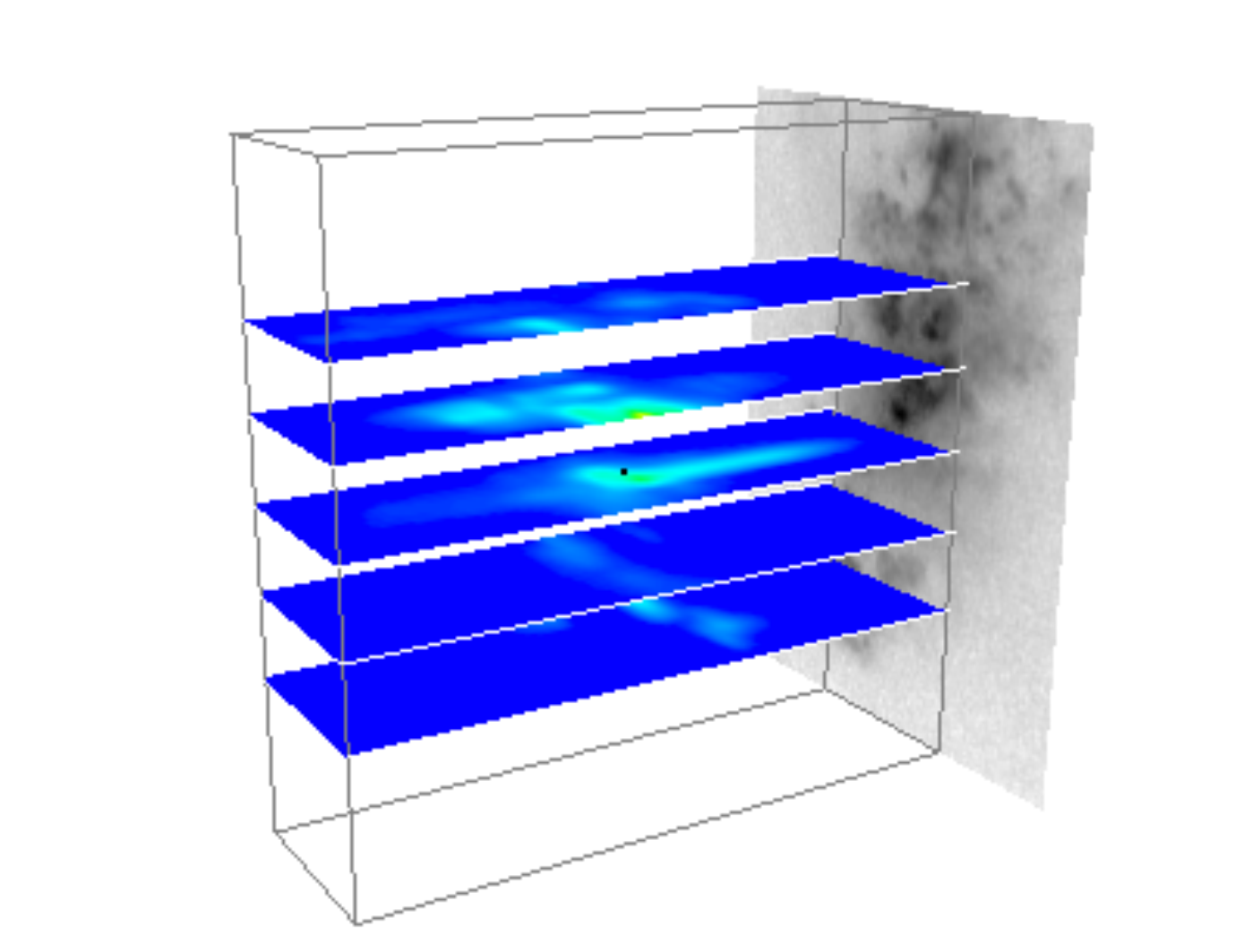} }}%
    
    \hspace{-1.5cm}
    \subfloat[sliced 2D images parallel to sky plane]{{\includegraphics[width=0.5\textwidth]{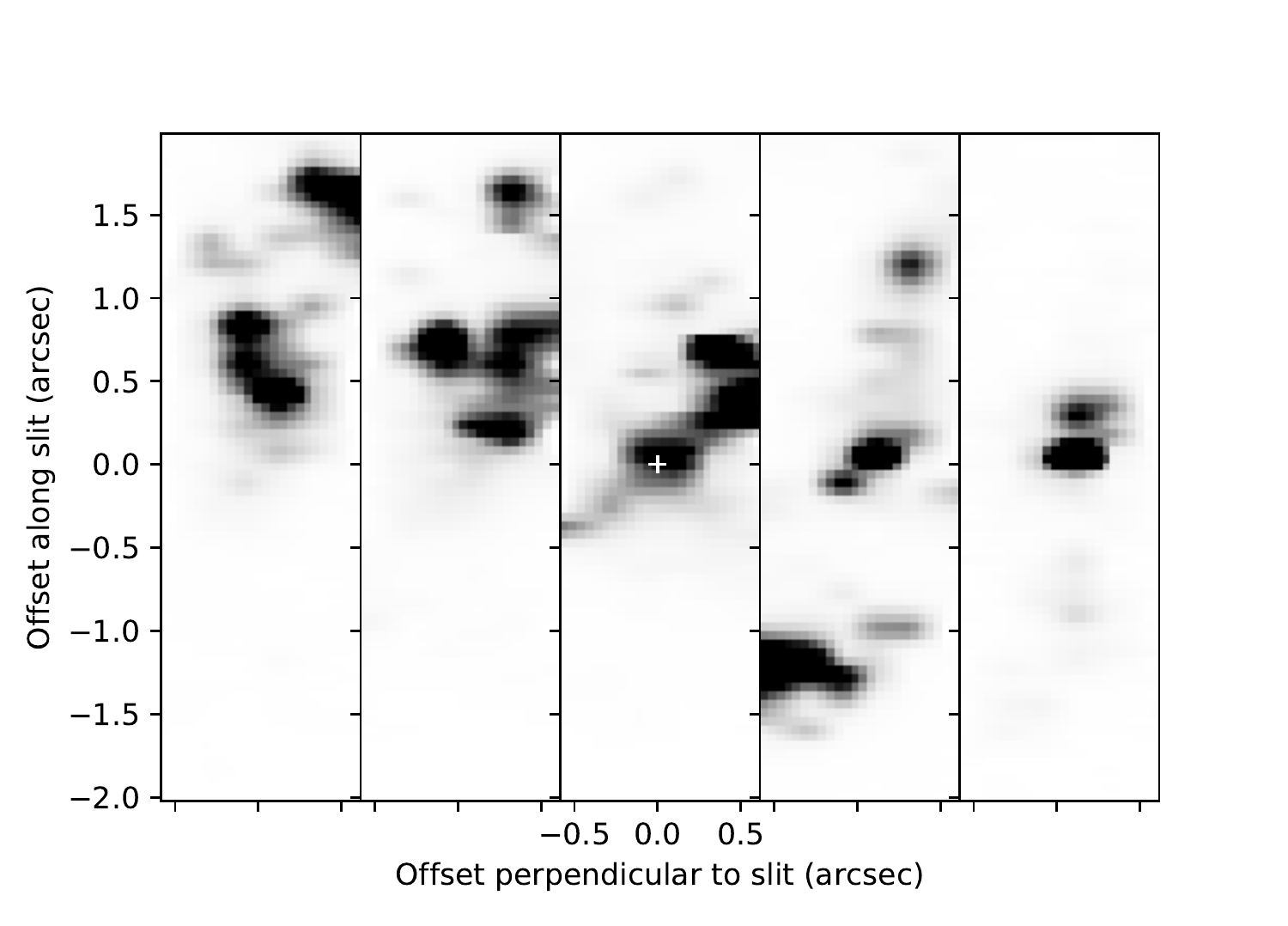} }}%
    %
    \hspace{+1.5cm}
    \subfloat[sliced 2D images along line of sight]{{\includegraphics[width=0.35\textwidth]{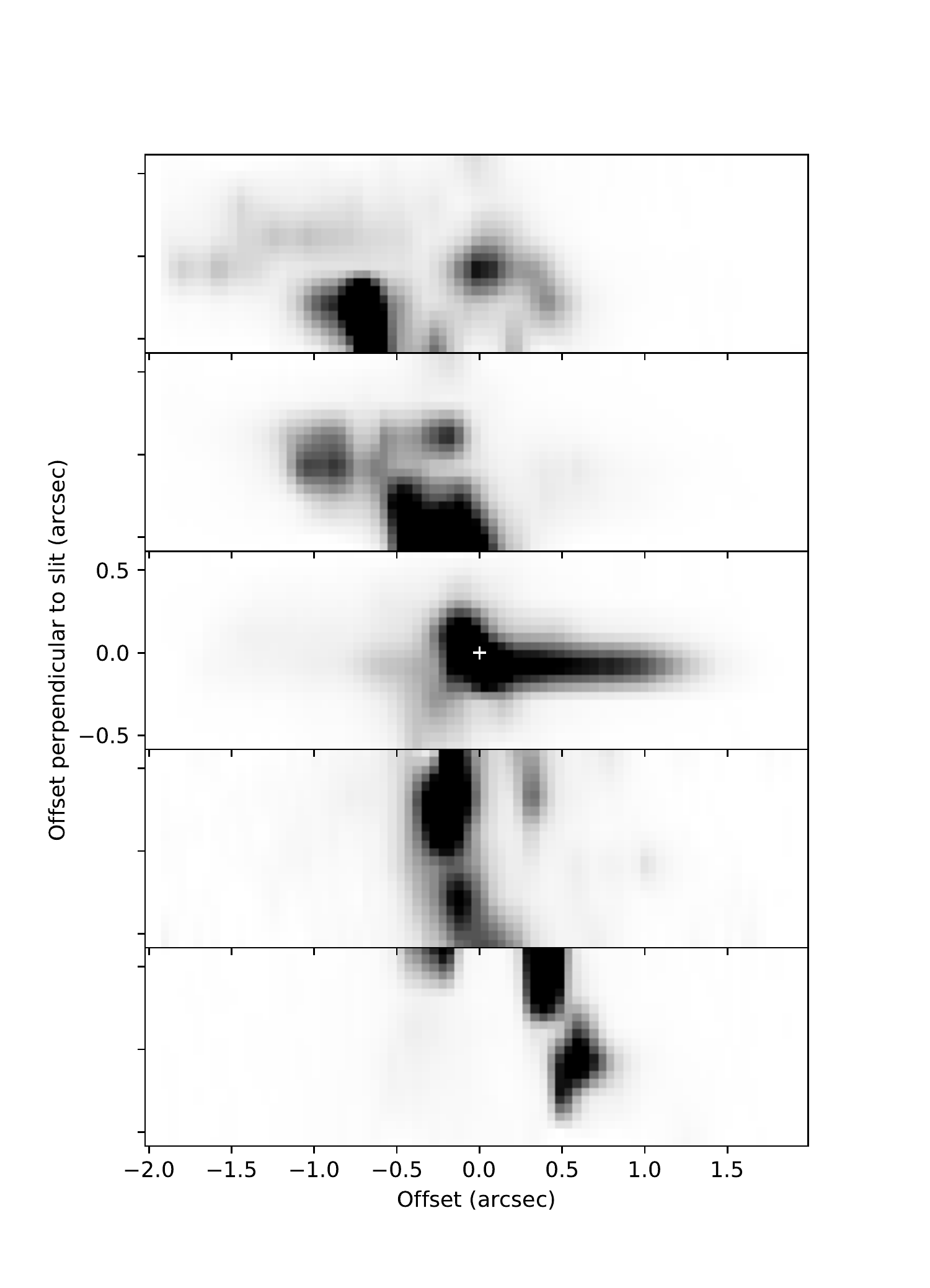} }}%
    \caption{Various 2D slices of the reconstructed 3D distribution. The position of the nucleus is shown as a black dot in panels (a) and b), and as a white plus sign in (c) and (d).}%
    \label{fig_slices}%
\end{figure*}

\section{Velocity-inversion mapping of the 3D flux distribution}\label{sec_method}

\subsection{3D mapping}
In this section, we describe how we reconstruct the 3D flux distribution from the resolved emission-line spectra, or radial velocities of the line-emitting gas clouds. We would need a few assumptions on the cloud velocities around the nucleus: (1) velocity vector $\bm{v}$ is parallel to the position vector $\bm{r}$ with respect to the nucleus ($\bm{v} \parallelslant \bm{r}$); (2) the velocity amplitude $v$ is determined as a certain function $f$ of the distance $r$ from the nucleus, i.e. $v = f(r)$. We also assume that we know the position of the nucleus projected on the sky plane which we designate as $x$-$y$ plane (see Fig.\ref{fig_vel_invert}). For Type 2 AGN, this is not necessarily the case, but for NGC1068, we do know the nucleus position quite accurately thanks to the HST imaging polarimetry data. 

If we have high spatial resolution integral field spectroscopic data of the line-emitting gas in the nuclear region, the observables for each line of sight are the line flux as a function of the radial velocity $\vrad$, and the projected distance $\rproj$ from the nucleus for that line of sight. The two quantities, $\vrad$ and $\rproj$, are related to the distance $z$ of the line emitting material from the sky plane as
\begin{equation}
\vrad = f(r) \frac{z}{r}
\end{equation}
where
\begin{equation}
r = \sqrt{\rproj^2 + z^2}.
\end{equation}
Therefore, in principle, we can convert $\vrad$ to $z$ for each line of sight, i.e., invert the observed radial velocity to the 3D position of the gas. In practice we might not have a unique solution of $z$ for a given $\vrad$ and $\rproj$, depending on the exact form of the velocity function $f(r)$. However, the inner velocity field has been inferred to have quite a simple form (\citealt{Crenshaw00,Das06}):
\begin{equation}
v = f(r) = k r.
\end{equation}
In this case, we simply have
\begin{equation}
\vrad = k z
\end{equation}
Thus the flux distribution in $xyv_r$ 3D space becomes equivalent to that in $xyz$ 3D space. A more general power-law form 
\begin{equation}
v = f(r) = k r^{\alpha}
\end{equation}
has also been investigated (\citealt{Das06}). In this case, we have
\begin{equation}
\vrad = k z (\rproj^2 + z^2)^{(\alpha -1)/2}.
\end{equation}
As long as the right hand side is a monotonic function of $z$, we can simply invert $\vrad$ to obtain $z$. Below we restrict the inversion mapping to the inner region, where the velocity field seems apparently linear, or the "accelerating region" as designated by \cite{Das06} which is about $\pm$2" along the slit from the nucleus, and implement this inversion process in this region only.

\subsection{Determination of the constant $k$}
\label{sec_method_k}

The mapping above does not specify the free parameter $k$, which sets the scale in $z$-direction with respect to that in the $x$-$y$ plane. Its approximate value can easily be obtained as $v_r / r_{\rm proj}$.  Since this is only a scaling in one direction, our analysis below is not affected by the detailed value of $k$. Nevertheless, we can determine $k$ in a relatively robust way.

A reasonable assumption would be that our line of sight is not special -- the spread of the flux distribution in $z$-direction is similar to that in $x$-$y$ plane. Here we propose to calculate $k$ from $\sigma_{v_r} = k \sigma_z$ where $\sigma_z$ is given as 
\begin{equation}
\sigma_z^2 = \frac{\sigma_x^2 + \sigma_y^2}{2}.
\label{eq_sigz}
\end{equation}
Here each $\sigma$ denotes the flux-weighted dispersion in each axis direction. This dispersion is calculated from
\begin{equation}
\sigma_x^2 = \frac{\sum (x-\bar{x})^2 \cdot F_{\rm OIII} (x,y,v_r)}{\sum F_{\rm OIII} (x,y,v_r)}
\end{equation}
and
\begin{equation}
\bar{x} = \frac{\sum x \cdot F_{\rm OIII} (x,y,v_r)}{\sum F_{\rm OIII} (x,y,v_r)}
\end{equation}
where $F_{\rm OIII}$ is the [OIII]5007 line flux distribution in $xyv_r$ space and the summation is taken over this space. The dispersions $\sigma_y$ and $\sigma_{v_r}$ are calculated in the same way, so that we can determine the constant $k$.

As for the uncertainty of $k$, we would postulate that $\sigma_z^2$ could be as large as $\sigma_x^2 + \sigma_y^2$, and at minimum it would be equal to $\sigma_x^2$ or $\sigma_y^2$, whichever smaller. Thus, $\sigma_z^2$ as determined by the above equation~\ref{eq_sigz} would have an uncertainty of roughly a factor of 2. Therefore, $\sigma_z$, and thus $k$, would roughly have an uncertainty of a factor of $\sqrt{2}$.

\begin{figure*}[ht!]
    \centering
    \subfloat[$v \propto r^{0.5}$]{{\includegraphics[width=0.3\textwidth]{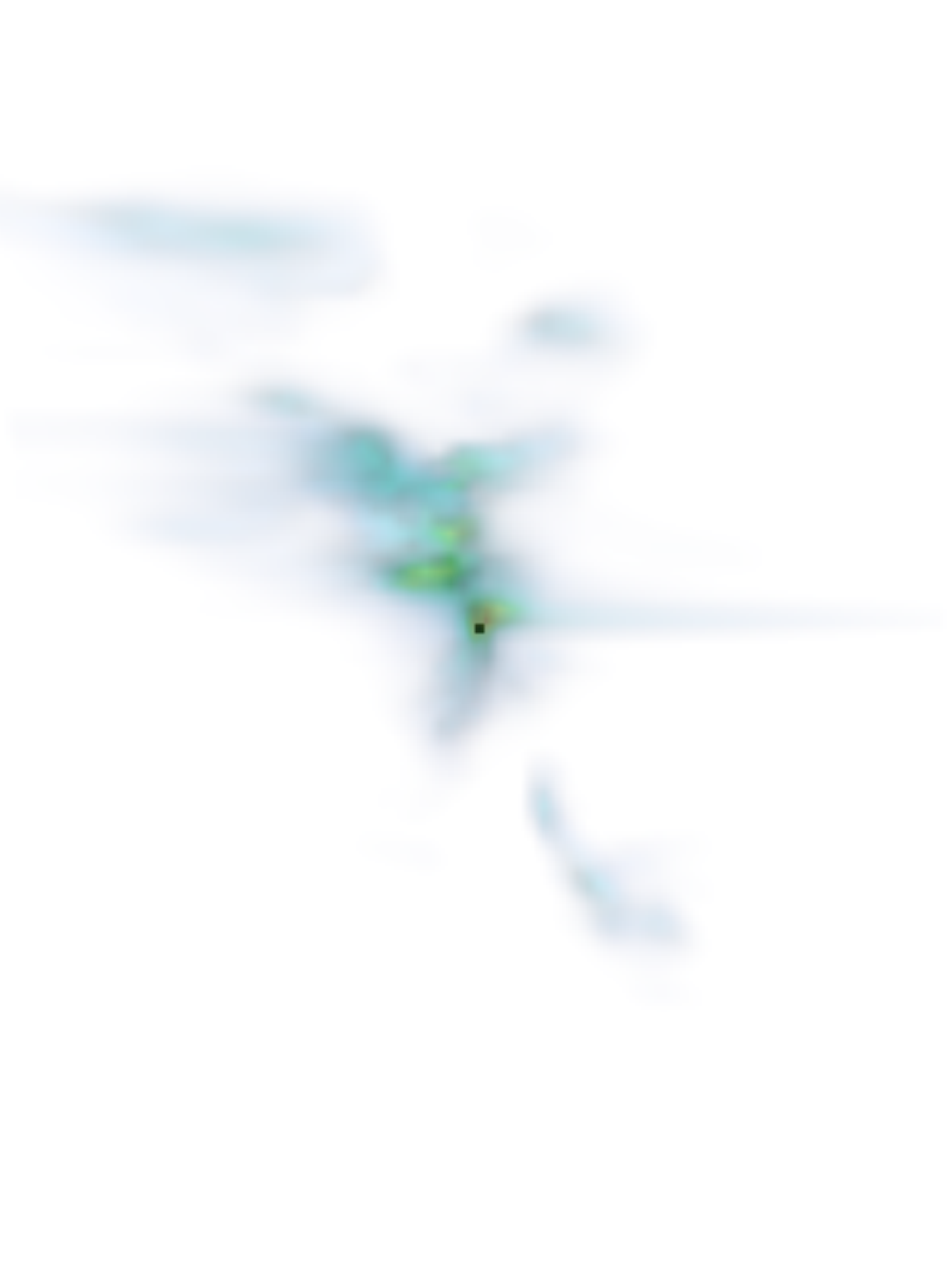} }}%
    \subfloat[$v \propto r^{1.0}$]{{\includegraphics[width=0.3\textwidth]{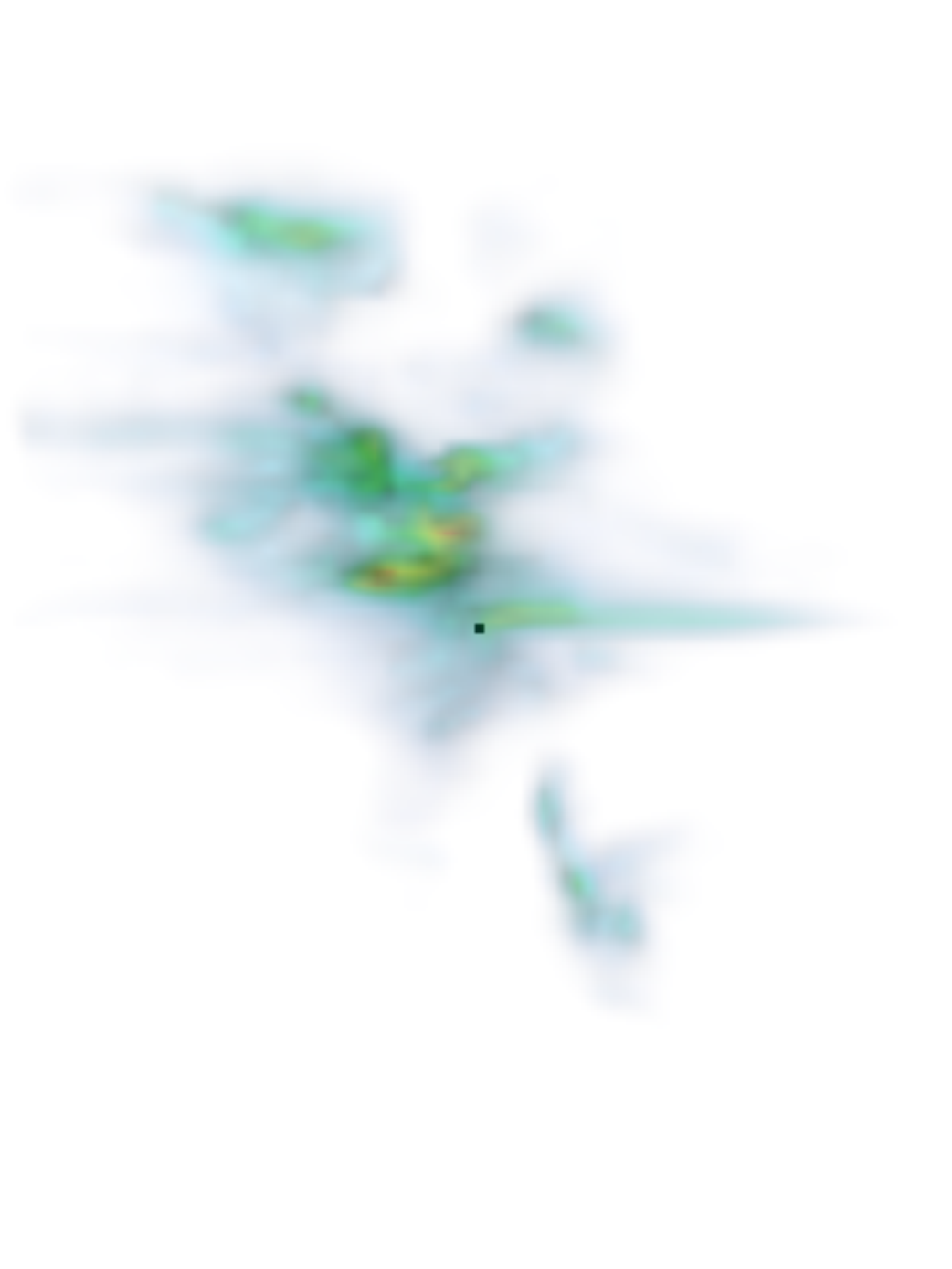} }}%
    \subfloat[$v \propto r^{1.5}$]{{\includegraphics[width=0.3\textwidth]{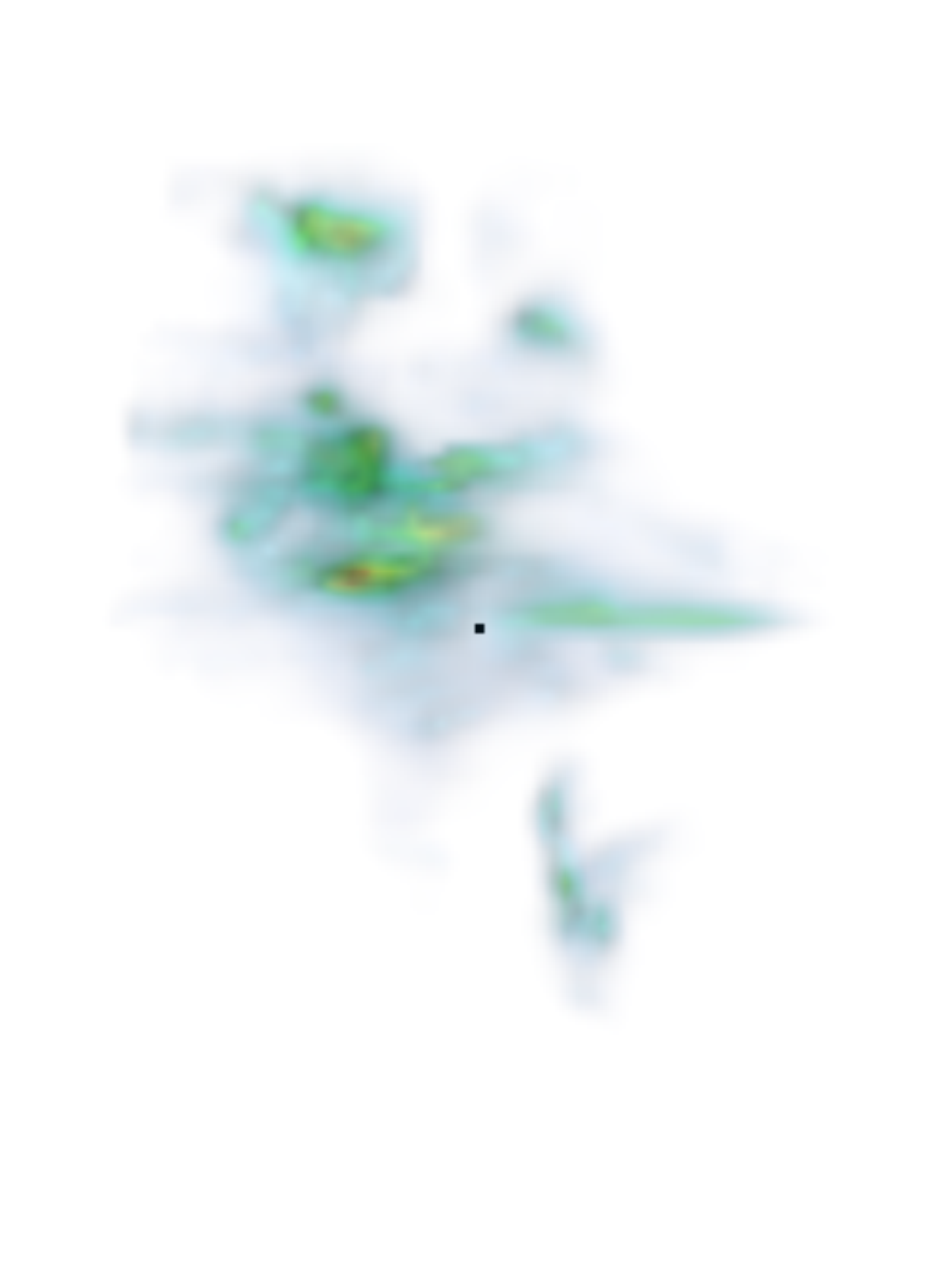} }}%
    \caption{Comparison of the reconstructed 3D distribution using different power-law indices in the velocity law. All representations are viewed from +$x$-axis direction, parallel to the sky plane.}
\label{fig_3D_comp}
\end{figure*}

\section{Reconstructed 3D view of the nuclear region}\label{sec_res}

In Figure~\ref{fig_3D}, we show the results of this 3D reconstruction for the linear velocity field case of $v=kr$, using the visualization software {\it Mayavi} \citep{ramachandran11}. As described in the previous section, for this linear velocity field, the "reconstructed" 3D distribution is essentially the original $xyv_r$ data cube of the [OIII]5007 line in shape (after continuum subtraction and [OIII]4959 line removal), except for the scaling in the $v_r$ direction. The 3D flux distribution is shown in a volumetric visualization in the left panel, while two slices of the same distribution, going through the position of the nucleus (i.e. the black hole optically hidden from direct view) indicated by a black dot, are shown in the right panel. We also show further sets of 2D cuts in Figure~\ref{fig_slices}. 


We have determined the constant $k$ to have the dispersion of the flux distribution in $z$-direction the same as that in the sky plane, as described in section \ref{sec_method_k}, where the value of $k$ resulted in 1.04$\times10^3$ km/s/arcsec, or 14.8 km/s/pc, for this data set. There is remaining uncertainty in $k$, but we note again that a different $k$ would only scale the 3D distribution in $z$-direction (with respect to the sky plane; Fig.\ref{fig_vel_invert}). The remaining uncertainty in $k$ would not leave considerable uncertainty in the overall 3D shape. 

A few clouds, the cloud B in the \cite{Evans91} notation in particular which is just north of the nucleus, have a shape quite spread in $z$-direction. This is due to the large velocity dispersion of the gas in these clouds. The velocity inversion mapping adopted here would not differentiate radial velocities from velocity dispersions, hence a large velocity dispersion produces an artifact on the 3D distribution. However, the velocity dispersions in most of the clouds seem to be of the same order of, or less than, the volumetric spread over the $x$-$y$ plane, and this issue does not affect the overall reconstructed distribution.

While somewhat messy, the distribution is certainly biconical in 3D, with the southern side rather extinct perhaps due to the absorption by the host galaxy disk. Even more importantly, the conical distribution seems roughly hollow. This is all consistent with the model proposed by \cite{Das06} and \cite{Crenshaw00} originally. Here we show this distribution in 3D directly from the data. We further discuss the implications of this outflowing, hollow structure in section \ref{sec_disc}.

In Figure~\ref{fig_3D_comp}, we show the 3D distribution reconstructed with different power-law indices of the velocity field. The detailed shape of course changes, but the overall distribution is not so sensitive to the choice of the index. Furthermore, we argue in section~\ref{sec_disc} that the linear velocity distribution (i.e. the $\alpha=1$ case) has more physical background and motivation.

\section{Discussion}\label{sec_disc}

\subsection{Comparison with previous modelling}

The biconical hollow geometry in outflow for NGC1068 at this 100 pc scale has been adopted and investigated in details by \cite{Das06}. The outflow geometry has originally been proposed by \cite{Crenshaw00} based on the HST/STIS single long-slit data as well as the conical morphology seen in earlier imaging data including HST images (\citealt{Evans91,Macchetto94}). The single long-slit data showed mainly two separate velocity components at each spatial location along the slit, and the data have been argued to be well explained if the biconical hollow geometry {\it and} the linear velocity field ($v=kr$) are both assumed. The latter assumption originates from the radial velocity being seemingly proportional to the projected distance from the nucleus. The multi-slit data have strengthened this idea, and \cite{Das06} have modeled the data based on this biconical hollow geometry and determined a few critical model parameters.

Our approach here is that, if we assume this linear velocity field, we can actually invert the radial velocity data to derive and reconstruct directly the 3D distribution of the line-emitting material. The only parameter we have to determine is the proportionality constant $k$, but this has only a scaling effect in one direction. We determined it to have the distribution 'round', i.e. having roughly the same spatial extent when viewed from different directions. This assumption approximately corresponds to the modelling by \cite{Das06} where the distribution is set to be axisymmetric. In fact, they obtained from the fit the value of the constant $k$ in the linear velocity law to be 14.3 km/s/pc (2000 km/s per 140 pc; from Table~2 in \citealt{Das06}), while we deduced essentially the same value (14.8 km/s/pc; sect.\ref{sec_res}). Our results support the overall biconical, hollow geometry proposed by \cite{Crenshaw00} and \cite{Das06}, thus confirm their modelling results. The reconstructed 3D distribution could give even more details and show the deviation from the simple geometry. However, we have to keep in mind that the reconstruction is based on the assumption of the velocity field, which we discuss in the next section.

\subsection{Physical interpretation of the velocity field}

Detailed investigations have been done on the interpretation of the seemingly linear velocity field (\citealt{Das07}). We could interpret the field as a given gas cloud being accelerated along the radial track from the nucleus. For a cloud with a given mass, a constant force (acceleration) would result in the cloud's velocity $v \propto r^{1/2}$ where $r$ is the distance from the nucleus. The case of the linear velocity $v \propto r$ would mean that the force $F$ is increasing in proportion to $r$ ($F \propto r$). If this accelerating force is the radiation pressure on dust grains, we would have to expand the cross section $S$ of the blob, keeping the whole blob still UV-optically-thick, (1) with $S \propto r^2$ to keep the force constant, or (2) with $S \propto r^3$ to have $F \propto r$. \cite{Das07} have investigated different cases of effective acceleration, concluding that the observed gradual acceleration of $v \propto r$ is quite difficult to achieve. Rather, the acceleration is quite possible in the innermost scale, and the cloud would keep the same speed over large distances if without a significant drag force from the ambient medium.

\subsection{ "AGN big bang" -- episodic acceleration}

Another way of interpreting this linear velocity field is that each cloud is actually moving at its own constant speed. All the clouds got accelerated in the innermost region, and are moving away from the nucleus at constant velocities, like the Hubble expansion. \cite{Das07} noted the velocity field as Hubble-flow-like, and this Hubble-flow interpretation has been advocated by \cite{Ozaki09}. In this case, the clouds are interpreted to have a constant velocity over the dynamical time $t_{\rm dyn}$ of $\sim r/v=1/k$. This would mean that an episodic acceleration, or an "AGN Big Bang", has occurred at the nucleus back in time of $1/k$. Since we have $k \sim 15$ km/s/pc for this inner region of NGC1068, the time scale $t_{\rm dyn}$ would roughly be $\sim 0.7 \times 10^5$ yr.

This explanation of the linear velocity field would give a credit to the 3D reconstructed distribution assuming $v \propto r$, and not the other power-law forms. In addition, this interpretation would be valid only to the region within a certain radius where the velocity could be kept approximately constant for a given cloud without a significant decelerating cause. We have limited our analysis to the region within 2 arcsec from the nucleus where the radial velocity seems proportional to the projected distance (see Fig.\ref{fig_2dspec}; called "accelerating region" by \citealt{Das06}). For the region outside, we would need a further consideration, but the clouds there could belong to a different episode of acceleration, further in the past.

\cite{Wang10} identified an extended soft-X-ray emission in NGC4151 at $\sim$10$^4$ light years (a few kpc) from the nucleus. Based on its high surface brightness, they infer that this could be due to an episodic outburst of the nucleus which occurred $\sim$10$^4$$-$10$^5$ years ago. NGC4151 has an NLR-scale outflow, spatially and spectrally resolved by HST (\citealt{Das05}), very similar to the outflow we discuss here, while NGC1068 does not seem to show such a large scale extended X-ray emission beyond kpc scale (\citealt{Ogle03,Kallman14}). It would be interesting to study different AGNs with spatially-resolved inner-NLR outflows systematically, in light of the scenario where the outflow is from an episodic outburst and in the resulting Big-Bang-like velocity field. Under this scenario, the estimation of the event time scale is relatively robust (the uncertainty corresponds to that in $k$, thus would roughly be a factor of $\sqrt{2}$), and the time scale might be correlated with the characteristics of the objects. We aim to implement these studies elsewhere.

\begin{figure}[t]
\vspace{-3mm}%
\hspace{-5mm}%
\includegraphics[width=10.5cm]{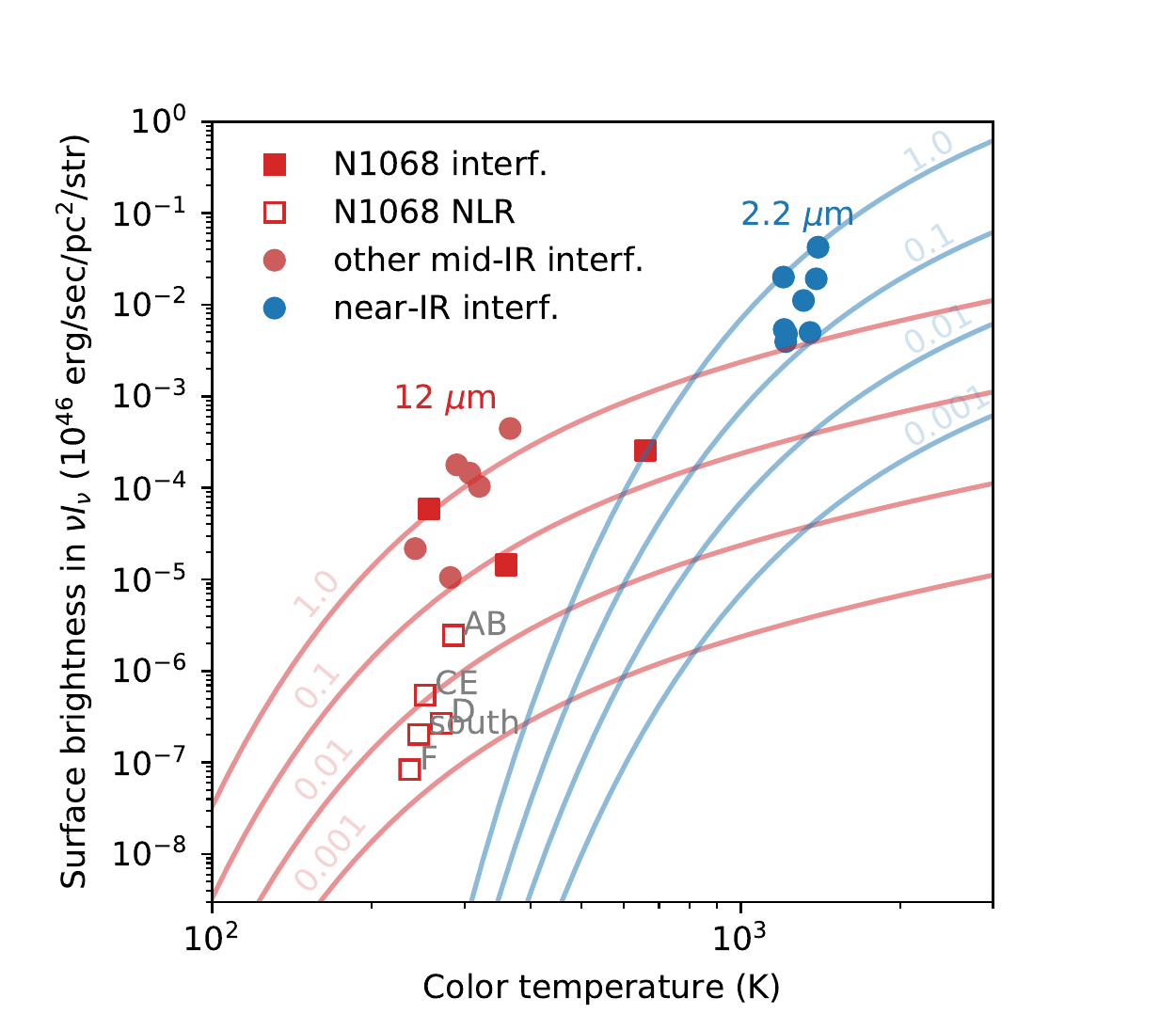}
\caption{
Surface brightness of the pc-scale region of NGC1068 (red filled squares; mid-IR interferometry data from \citealt{LopezGonzaga14}) and several Type 1 AGNs with IR interferometric and color temperature measurements (\citealt{Kishimoto11,Kishimoto11MIDI,Burtscher13,Dexter19}). The mid-IR data for the resolved clouds in NGC1068, 3D-reconstructed here, are shown in open squares (clouds A/B, C/E, D, F, south; data from \citealt{Bock00}). Red curves show the surface brightness of the gray body at 12 $\mu$m with emissivities from 1 (i.e. black body) to 0.001. The blue curves are the same but at 2.2 $\mu$m. 
}
\label{fig_emissivity}
\end{figure}

\begin{figure*}[t]
\centering
\includegraphics[width=\textwidth]{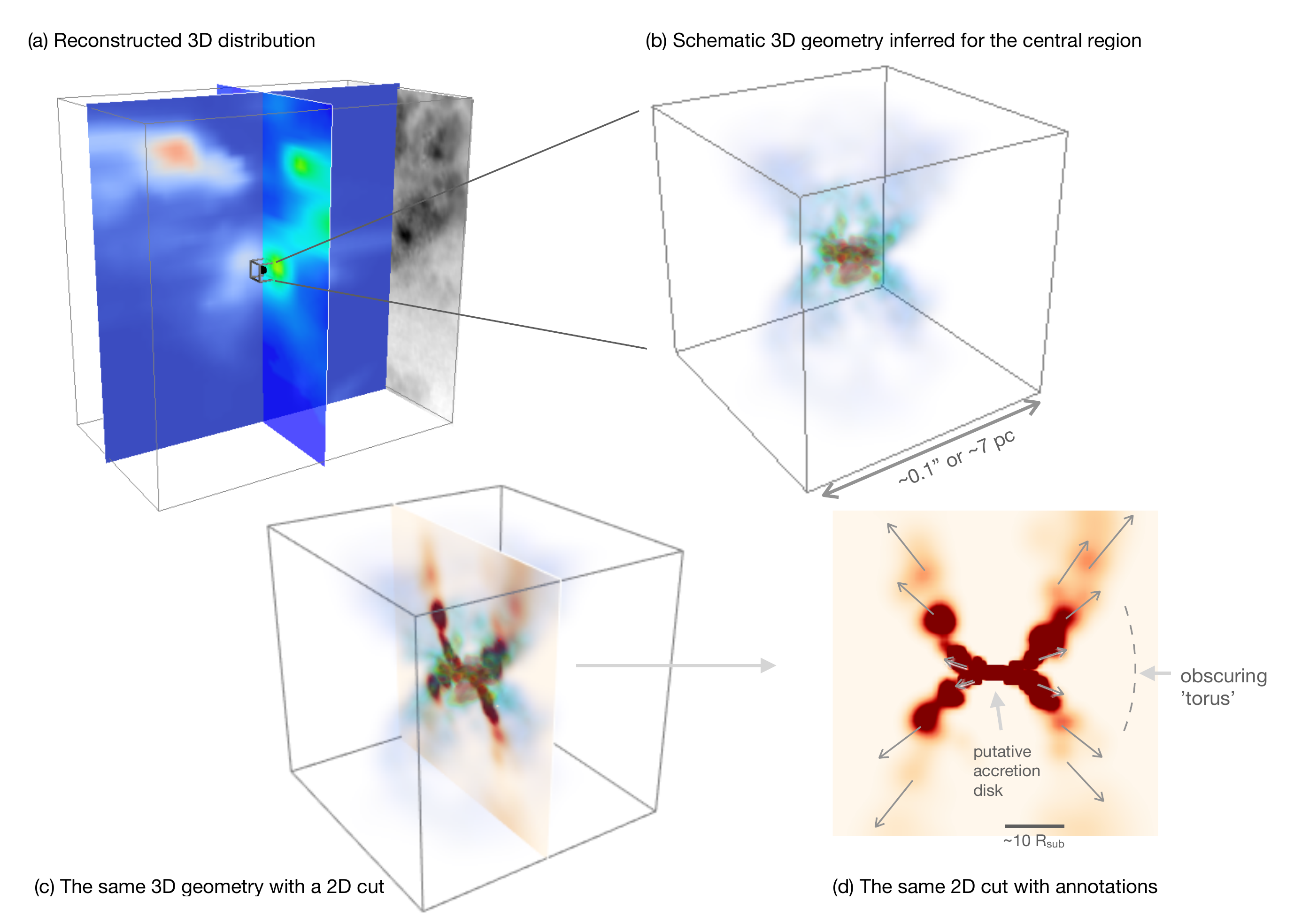}
\caption{The central part of the 3D structure of Fig.\ref{fig_3D}b is shown in panel (a) with a $0."1$ central cube ($\sim$7 pc), indicating the approximate size for the schematic 3D illustration shown in panel (b). The same 3D illustration is shown in panel (c) with a central 2D cut, which is shown with annotations in panel (d).}
\label{fig_geometry}
\end{figure*}

\subsection{Comparison with the interferometrically resolved inner structure}
\label{sec_disc_comp}

It is quite instructive to compare the conical and hollow 3D distribution at this 100 pc scale with the structure observed for a much inner region at a pc scale. We summarize here four important observational aspects below, and discuss their implications in the next section.

(1) First, recent IR interferometry has shown that the mid-IR emission at the inner pc scale, or 10s of dust sublimation radii $\Rsub$, is polar elongated, rather than equatorially elongated (\citealt{Hoenig12,Hoenig13,Tristram14,LopezGonzaga14,LopezGonzaga16}). Therefore the inner pc-scale structure is extended toward the conical and hollow distribution in the NLR at an arcsec or 100 pc scale that we discuss here. A mid-IR polar elongation is seen in many nearby AGNs with sub-arcsec mid-IR imaging (\citealt{Asmus16}).

(2) On the other hand, when the effective resolution is high enough to resolve even an inner scale of several $\Rsub$ in the mid-IR, which is the case for the two Type 2 objects NGC1068 and Circinus galaxy, we see an equatorially elongated structure, in addition to the slightly outer polar structure (\citealt{LopezGonzaga14,Tristram14}).

(3) Quite importantly, the measured size and flux indicate a mid-IR emissivity of sub-unity, i.e. $\sim$0.1 -- 1. We summarize this in Figure~\ref{fig_emissivity}. The flux and emitting region size measured at a certain wavelength give us the estimation of the surface brightness. With a further estimation of the color temperature from the spectral shape, we can estimate the emissivity, i.e. the ratio of the surface brightness of the region to that of the black-body emission at that temperature. The three red filled squares in Figure~\ref{fig_emissivity} are the  mid-IR interferometric measurements for NGC1068, modelled with three gaussian components, each giving a set of size, flux, and temperature (\citealt{LopezGonzaga14}). We also collected mid-IR interferometric size measurements with temperature estimation in the literature, as well as similar near-IR interferometric measurements, all shown in filled circles in Figure~\ref{fig_emissivity} (\citealt{Kishimoto11,Kishimoto11MIDI,Burtscher13,Dexter19}). The data are plotted on the grids of gray-body surface brightnesses as a function of temperature with emissivities from 1 (i.e. black body) to 0.001 at 12 $\mu$m (red) and 2.2 $\mu$m (blue). The data points indicate that the nuclear IR emission is quite close to the black body surface brightness -- some of the 12 $\mu$m points are above the black-body curve which means that the temperature of the emitting material is probably somewhat higher than the color temperature. But in any case, the emissivity is all sub-unity.

(4) In comparison to these emissivity estimates for the inner pc-scale structure, we can estimate the emissivity of the 100-pc-scale HST-resolved clouds in NGC1068 using the resolved mid-IR flux of the clouds as quantified by \cite{Bock00}. They measured the spatially-resolved mid-IR SED for five different clouds of 0."2 in diameter, corresponding in the notation of \cite{Evans91} to the clouds A/B, C/E, D, F, and the clouds just south from the nucleus. These five measurements are indicated accordingly in Figure~\ref{fig_emissivity} with open squares. The emissivity of these discrete clouds is $\sim$0.001 -- 0.01, much lower than that of the inner structure in general, and strikingly, the loci of these clouds on Figure~\ref{fig_emissivity} look quite continuous from the interferometrically resolved structure. Therefore these two structures should be closely connected, as we discuss in the next section.

\subsection{Obscuration with outflow: "flaring windy torus"}

The emissivity of sub-unity found for the inner pc-scale structure is in fact the value expected for the surface of the UV-optically-thick clouds directly illuminated by the nucleus and observed in the IR (see e.g. Fig.3 of \citealt{Hoenig10model}). 
This implies that these inner (interferometrically resolved) {\it polar} clouds at pc-scale are {\it participating in the obscuration of the nucleus}, i.e. these must be a part of the AGN "torus". Then, to avoid polar obscuration (to have the Type-1 line-of-sight clear), we would need to distribute these UV-optically-thick clouds in a hollow geometry, and likely in a flaring configuration in order to have each cloud directly illuminated by the nucleus. For the case of NGC1068, we indicate in Figure~\ref{fig_geometry}a the approximate spatial scale discussed here as the inner box of $0.1"$ side, corresponding to $\sim$7 pc. This is a region of $\sim$30 $\Rsub$ in radius. The schematic 3D illustration of the distribution of the ionized gas and high temperature dust ($\gtrsim$300 K) inferred here for this inner region is shown in Figure~\ref{fig_geometry}b, with a 2D cut shown in Figure~\ref{fig_geometry}c and d. This is quite similar to the configuration described by \cite{Hoenig17} and \cite{Hoenig19}, but the flaring geometry proposed here has both the inner equatorial elongation and outer polar elongation in a single component.

The conical and hollow 100-pc-scale distribution, 3D-reconstructed here, would give further support for the inferred inner structure illustrated in Figure~\ref{fig_geometry}b-d, which is also conical and hollow. We would infer that these two structures are smoothly connected, which is consistent with the continuous distribution in emissivity on Figure~\ref{fig_emissivity}. The inferred flaring configuration is also consistent with the outer-polar and inner-equatorial elongation (combination of the points 1 and 2 in section~\ref{sec_disc_comp}).

The inferred hollow/flaring inner geometry does not directly have any kinematic information yet. However, since the 100-pc-scale outflow is very likely to be connected to this inner pc-scale structure, the inner one must be in outflow as well. We would then suggest that the AGN "torus", the obscuring structure, would indeed be the inner part of this hollow conical outflow. Thus it is an obscuring {\it and} outflowing structure. The HST-resolved high velocity clouds would probably be the outer part of this flaring windy torus, expanded at large distances from the nucleus and become optically-thin, consistent with the mid-IR emissivity of 0.01-0.001 (point 4 in section~\ref{sec_disc_comp}).

The inferred inner geometry could actually be consistent with 
the reconstructed image of the pc-scale region of NGC1068, which has recently been obtained with VLTI/GRAVITY in the near-IR (\citealt{Pfuhl20}). However,  the comparison might not be straightforward due to the possible significant extinction in the near-IR for this Type 2 object.

\subsection{Relations to models and sub-mm observations}

This observational picture seems consistent with the theoretical outflow models developed by e.g. \cite{Koenigl94} and \cite{Wada12,Wada15} which posit the inner part of the outflow as the obscuring structure. Here we argued that such a conical, hollow outflow is observationally seen (i.e. can be reconstructed) at $\sim$100 pc scale, and can be well inferred at $\sim$pc scale as its inward extension based on the IR interferometric observations.

This structure traced by the ionized gas and high-temperature dust grains probably has quite a good correspondence to that traced by molecular gas observed in sub-mm high-resolution maps for NGC1068. In the inner UV-optically-thick part of the outflow, the illuminated "bright" side has ionized gas and high-temperature dust, whereas the shadowed, "dark" side of each of these clouds has molecular gas and low-temperature dust (see Table~\ref{tab_prop} for summary). 

ALMA has now the resolution to spatially resolve the dark side of the hollow conical distribution (Fig.8d), and we do see an apparently corresponding, X-shaped morphology in sub-mm molecular lines and continuum: CO ($J$=6$-$5) at 4 pc (0.06") resolution \citep{GarciaBurillo16}; 256GHz continuum at 1.4 pc (0.02") resolution \citep{Impellizzeri19}; HCO$^+$ (4$-$3) at 2 pc (0.03") resolution \citep{GarciaBurillo19}. The spatial distribution of the high velocity CO (6$-$5) line emission peaks (\citealt{Gallimore16}) could also correspond to the dark sides of the outflow clouds. A part of HCN (3$-$2) line emission and absorption also shows outflow features in radial velocities at the high enough spatial resolution (\citealt{Impellizzeri19}). We also note that outflow velocities of these molecular gas ($\sim$100 km/s) seem roughly consistent with the inward extension of the linear velocity law discussed here (e.g. 75 km/s at 5 pc for $k$=15 km/s/pc; sect.\ref{sec_res}).

One complication in sub-mm seems to be that we observe both an outflow and a rotating gas (\citealt{Imanishi18,Impellizzeri19,GarciaBurillo19}) -- the latter is likely in the equatorial plane and perhaps fueling the nucleus, thus could be called a 'fueling disk' component (Table~\ref{tab_prop}). Disentangling the two has seen a substantial progress with better spatial resolutions. As discussed by \cite{Hoenig19}, the results of high-angular-resolution observations in the infrared and sub-mm can be understood in a unifying way if we take into account the two-component nature of the sub-mm emitting material.

\begin{table}
\caption{Inferred properties of various components}
\label{tab_prop}
\begin{tabular}{lcccc}
\hline
\hline
 & \multicolumn{3}{c}{outflow} & \multirow{2}{*}{\parbox{1.5cm}{\centering fueling disk}} \\
\cline{2-4}
\cline{2-4}
 & NLR & \multicolumn{2}{c}{obscuring torus} &  \\
\cline{3-4}
 &     & bright side & dark side & \\
\hline
Scale & $\sim$100 pc & \multicolumn{2}{c}{$\sim$1 pc} & $\sim$10 pc \\
UV-thickness  & opt-thin & \multicolumn{2}{c}{opt-thick} \\
Location      & polar & \multicolumn{2}{c}{polar/equatorial} & equatorial \\
\multirow{2}{*}{\parbox{1.7cm}{Ionized gas / high-T dust}} & \multirow{2}{*}{$\Circle$} & \multirow{2}{*}{$\Circle$} &  
\multirow{2}{*}{--} & \multirow{2}{*}{--} \\
& & & & \\
\multirow{2}{*}{\parbox{1.9cm}{Molecular gas / low-T dust}} &  
\multirow{2}{*}{--} & \multirow{2}{*}{--} &  
\multirow{2}{*}{$\Circle$} & \multirow{2}{*}{$\Circle$}\\
& & & & \\
\hline
\end{tabular}
\end{table}

\section{Summary and conclusions}

Two-dimensional spectral imaging data have become quite abundant over the recent years. We have shown that the 2D radial velocity map from such data directly gives us the distribution of emission line flux in three-dimensional space, if we assume a certain velocity field. We used archival HST/STIS multi-slit data for the prototypical Seyfert 2 galaxy NGC1068, and reconstructed the 3D distribution of the line flux directly from the high spatial resolution velocity map.

Among various velocity laws, the linear velocity field can in principle be physically understood with the scenario in which all the clouds were accelerated simultaneously in a very short time scale in the innermost region. The clouds have been moving away from the nucleus with a constant speed, leading to the radial distance of each cloud from the nucleus being proportional to the velocity of that cloud. This "AGN big-bang" event, inferred to have occurred $\sim$$10^5$ years ago, could be an episodic one in the history of the object. 

The 3D distribution of the emission-line flux reconstructed with this linear velocity law shows a roughly hollow bi-conical structure at a 100 pc scale. Combined with the recent IR interferometric pc-scale results, we infer that this hollow conical outflow extends from the inner pc scale to the 100 pc scale. Since the emissivity of the inner pc-scale clouds is consistent with being UV-optically-thick, those outflowing clouds are considered to be obscuring the nucleus at the pc scale. Thus the AGN 'torus' is argued to be the obscuring {\it and} outflowing structure. The corresponding, shadowed side of this inferred pc-scale structure now seems to be observed in sub-mm.

The outflow, or the instant acceleration in the innermost region, might have been caused by the radiation pressure on dust grains. If the illuminating radiation is anisotropic, being polar strong, the observed polar hollow structure might originate from such anisotropic radiation from the nucleus. In fact, such anisotropy could in principle be constrained from the 3D distribution obtained here. We plan to apply our analysis to many other AGNs with high spatial resolution velocity maps in a future publication.

\acknowledgements

We thank the referee for providing very constructive comments which substantially improved the paper. This work was supported by JSPS grants 16H05731 / 20K04029, and the grant E1906 through Kyoto Sangyo University.

\end{document}